\documentclass[12pt, letterpaper]{article}
\usepackage[pdftex]{graphicx}

\usepackage{subcaption}
\usepackage{multirow}
\usepackage{flushend}
\usepackage[cmex10]{amsmath}
\usepackage{color}
\usepackage{fancyhdr}

\usepackage{multicol}
\usepackage{authblk}

\begin{document}

%
\title{SAFFRON: A Semi-Automated Framework for Software Requirements Prioritization}


\author[1]{	Syed Ali Asif}
\author[1]{Zarif Masud}
\author[2]{Rubaida Easmin}
\author[1]{Alim Ul Gias}
\affil[1]{Institute of Information Technology,	University of Dhaka, Bangladesh}
\affil[2]{Department of Software Engineering, Daffodil International University, Bangladesh}

\date{}



%


\maketitle

\begin{abstract}
Due to dynamic nature of current software development methods, changes in requirements are embraced and given proper consideration. However, this triggers the rank reversal problem which involves re-prioritizing requirements based on stakeholders' feedback. It incurs significant cost because of time elapsed in large number of human interactions. To solve this issue, a Semi-Automated Framework for soFtware Requirements priOritizatioN (SAFFRON) is presented in this paper. For a particular requirement, SAFFRON predicts appropriate stakeholders' ratings to reduce human interactions. Initially, item-item collaborative filtering is utilized to estimate similarity between new and previously elicited requirements. Using this similarity, stakeholders who are most likely to rate requirements are determined. Afterwards, collaborative filtering based on latent factor model is used to predict ratings of those stakeholders. The proposed approach is implemented and tested on RALIC dataset. The results illustrate consistent correlation, similar to state of the art approaches, with the ground truth. In addition, SAFFRON requires 13.5-27\% less human interaction for re-prioritizing requirements.
\end{abstract}


%

\section{Introduction}
\label{sec:intro}
In software development, projects have more candidate requirements than can be implemented within budget and time constraints. Requirements Prioritization (RP) \cite{sher2014requirements} is essential to select which requirements need to be implemented before the others. RP forms the basis for product and market planning and thus plays a critical role in determining budget and expenses of the project \cite{aurum2003fundamental}. RP can be incorporated later in these decision processes but it will exceedingly  increase the project cost \cite{boehm1984software}. It is thus cost-effective to have a prioritized list of requirements early on that is accurate and best serves different stakeholders' needs \cite{nuseibeh2000requirements}. This can save cost, decrease time for product development by ensuring proper plan and also help in finding requirement defects. 

Requirements Prioritization has, however, been proven to be extremely challenging and one of the biggest issues is scalability. In large scale projects, the number of stakeholders is vast. These stakeholders are split in divisions and organizations. Each of them can have different needs which may create conflict in deciding which requirements need to be prioritized \cite{nuseibeh2000requirements,lim2011social}. Besides, there are various complexities such as inadequate budget, unskilled programmers, lack of resources and time. These complexities increase the need for human interaction that becomes infeasible when stakeholder size is too large. One major challenge that arises from the scalability issues is the rank reversal problem \cite{kukreja2013value}. 

Rank reversal means updating the ranking of the prioritized requirements when a new requirement is added or deleted or an old requirement is changed. As software development is an iterative process, requirements are identified during different phases such as designing, analysis or problem solving. Requirements can also change through client feedback. Thus the rank reversal problem is inevitable. It is particularly challenging for large scale projects with large number of stakeholders since such projects would have more volatile requirements that are subject to change. It is essential to take such changes into consideration when prioritizing requirements \cite{kukreja2013value,achimugu2014preference,achimugu2014clustering}.

Due to the necessity and benefits of RP, it has long been an active area of research. Researchers discussed about several stakeholder prioritization concepts for requirements prioritization: exploring collaboration \cite{damian2007collaboration}, risks of stakeholders' being negatively effected by project outcome \cite{woolridge2007stakeholder}, pairwise comparison \cite{karlsson1996software}, etc. The authors in \cite{kukreja2012selecting} used House of Quality (HoQ) framework \cite{hauser1988house} for comparative analysis of 17 requirements prioritization frameworks but none of these frameworks addressed the rank reversal problem. The authors in \cite{achimugu2014preference} used k-means algorithm  to solve rank reversal in requirements prioritization but failed to account for the stakeholder prioritization. Authors in  \cite{lim2011social, lim2011stakesource2, lim2012stakerare} one-by-one addressed problems like prioritizing stakeholders, identifying appropriate requirements and methods of prioritization however, they did not considered rank reversal problem. 

In this paper we propose a Semi-Automated Framework for soFtware Requirements priOritizatioN (SAFFRON) that addresses the rank reversal problem in requirement prioritization. Our proposed approach uses collaborative filtering techniques to resolve the rank reversal issues and decrease number of interactions with stakeholders. To the best of our knowledge, there exists no approach that has considered predictive models such as collaborative filtering to address the rank reversal issue. SAFFRON applies item based collaborative filtering (based on Pearson Correlation Coefficient) to determine similarities among new and already existing requirements. These similarities are later used to determine users who are highly likely to rate the new requirements. Model based collaborative filtering, which uses latent factor models \cite{agarwal2009regression} and gradient descent \cite{burges2005learning}, is then used for predicting ratings of the suggested stakeholders.

We implemented our proposed framework and compared the results against Ground truth and StakeRare \cite{lim2012stakerare} approach. It has been shown that proposed approach reduces human interaction by 13.5-27\% by maintaining strong ranking correlation with Ground Truth. The approach thus solves the rank reversal problem of requirements prioritization. Moreover, by reducing the human interactions, the approach is proven to be more scalable than StakeRare while yielding similar correlation with the ground truth.

Rest of the paper is organized as follows: Section \ref{sec:rel} covers related work on RP domain. Section \ref{sec:method} contains the proposed methodology. Section \ref{sec:exp} explained the experimental settings and section \ref{sec:result} discusses the results obtained from the experiment. The paper is concluded in section \ref{sec:conc} with future research directions. 

\section{Related Work}
\label{sec:rel}
Requirements prioritization is given importance by researchers since it helps in planning software releases in the scenario where all the requirements cannot be implemented in first release due to insufficient time and budget \cite{achimugu2014clustering}. Prioritization also enhances software testing by reducing the probability of generating ineffective test cases based on imprecise requirements. Researchers have focused on some necessary tasks for requirement prioritization - determining and classifying requirements, prioritizing stakeholders and selecting proper frameworks \cite{lim2012stakerare, lim2013using, achimugu2014web, achimugu2014adaptive}.

Authors in \cite{damian2007collaboration} studied the impact of distance in collaboration within social networks of stakeholders. The authors in \cite{woolridge2007stakeholder} recommended to consider risks of negatively effecting the stakeholders' during the prioritization process. Pairwise comparison and numeral assignment based strategies were used in \cite{karlsson1996software} to prioritize requirements of the project. Mitchell et al. \cite{mitchell1997toward} proposed a searching method for identification of stakeholders and their links. Authors in \cite{lim2010stakesource} automated stakeholder analysis by using crowd-sourcing approaches and prioritized stakeholders using Betweenness Centrality, Closeness Centrality and Page Rank Algorithm. 

In \cite{thakurta2013framework}, the authors presented quantitative framework for prioritizing nonfunctional requirements by using scenario-based approach. However, this approach fail to incorporate new requirements or change of existing ones and the evaluation suffered from validation issues. The research stated at \cite{achimugu2014web} introduced a multi-criteria decision making system- `Requirements Prioritizer' to prioritize requirements from any location. The system, while scalable and addressed the rank reversal issues persistent in techniques \cite{berander2005requirements} such as AHP, bubble sort, case base rank etc., had one major shortcoming. The approach did not prioritize or categorize stakeholders based on different requirement knowledge.  

The authors in \cite{achimugu2014adaptive} used Fuzzy multi-criteria decision-making (FMCDM) method to effectively deal with the inherent imprecision, vagueness and ambiguity associated with human decision making process in RP. Questionnaires to collect relative ranks from stakeholders were used to prioritize requirements. This approach did not prioritize stakeholders and also failed to take dependencies in requirements into account. Moreover, there could be assessment bias in the results of this approach.

In \cite{achimugu2014preference}, the authors supported stakeholder prioritization by ranking requirements based on the weight of their attributes provided by the relevant stakeholders. All the requirements must be mutually independent. This proposed approach deals with rank reversal and dependency issues. But the method of collecting requirement weight did not consider budget and time constraints.

K-means algorithm is used in \cite{achimugu2014clustering} to resolve rank reversal problem of large scale software prioritization. Multiple criteria are used to form clusters. The clusters were prioritized based on weights. However, the approach did not prioritizes stakeholders, used ambiguous methods to gain weights of requirements and did not handle dependencies in requirement prioritization. 

Lim et al. \cite{lim2011social} prioritized stakeholders using `StakeNet'- a social networking tool. This tool obtained recommendation of stakeholders from each stakeholder in the system through interviews. They extended this work to `StakeSource2.0' \cite{lim2011stakesource2}, which prioritizes requirements and stakeholders by means of social networking and collaborative filtering. Their work also highlighted stakeholders conflict and proposed recommending requirements to applicable stakeholders. However, rank reversal was not considered in either of these approaches.

The authors of \cite{lim2011stakesource2} also proposed `StakeRare' \cite{lim2012stakerare} which used social networks and collaborative filtering for large scale requirements. The paper addressed three problems for large scale projects: information overload, inadequate stakeholder input, and biased prioritization of requirements. The authors collected stakeholder list by eliciting requirements and deriving influence of stakeholders using interviews and the importance of each requirement was determined. From that, requirements were analyzed and a list of prioritized requirements were generated. Although the method performed well compared to other existing methods, it did not cover rank reversal problem. 

Review of state of the art framework for requirements prioritization illustrates that issues regarding rank reversal are not fully addressed in most of the approaches. Moreover, the approaches considering rank reversal suffers from several problems such as lack of stakeholder prioritization and computational complexity. In our knowledge, current approaches did not emphasize on reducing human interactions necessary for prioritizing new requirements. Depending solely on the feedback from stakeholders for prioritization, will increase time and cost needed for the process and introduce scalability issues.

\begin{figure}[h]
	\centering
	\includegraphics[width=\textwidth]{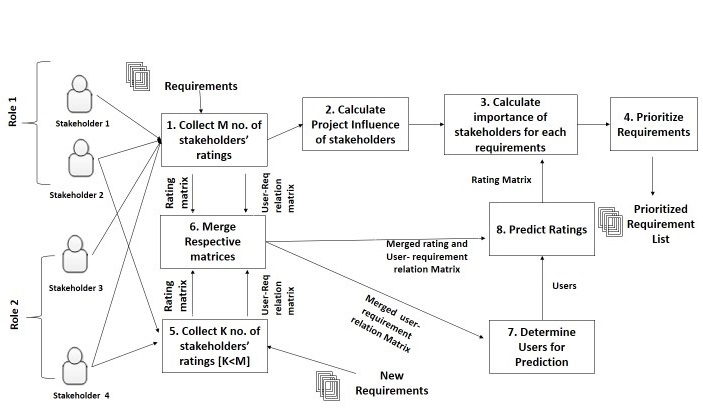}
	\caption{Overview of SAFFRON Framework}
	\label{fig_saffronFramework}
\end{figure}

\section{The SAFFRON Framework}
\label{sec:method}
This paper proposes a framework named SAFFRON which reduces human interaction while updating the ranks of the prioritized list after incorporating new requirements to already elicited and ranked requirements. By reducing the human interactions, it makes itself more suitable for large scale projects. SAFFRON consists of eight steps: initial collection of stakeholder's ratings, calculating project influence of stakeholders based on their roles and individual influences, computing importance of each stakeholders for every requirements from ratings and project influence, prioritizing requirements based on the total importance, for new requirements collecting ratings from a subset of stakeholders, merging both rating matrices of previous requirements and new requirements and deriving user-requirement relation matrix from merged rating matrix, determining users for whom to predict ratings, predict ratings using collaborative filtering. Lastly, new rating matrix with predicted ratings will be used as an input for step 3 and updated prioritized requirements list will be obtained. The first four steps of the framework are proposed by the authors in \cite{lim2012stakerare}. An overview of the whole framework is sketched in Figure \ref{fig_saffronFramework}.

The architecture can be divided into two separate parts. One part is concerned with prioritizing elicited requirements and the other part intends to solve rank reversal problem caused by new requirements.  Prioritizing elicited requirements follows these steps:
\begin{itemize}
	\item At first the requirements relevant to the project and its ratings will be elicited from the stakeholders using human interaction.
	\item Then the approach described in StakeRare \cite{berander2005requirements} will be applied. The stakeholders will be prioritized using the ratings provided by other stakeholders.
	\item After that each stakeholders' influence on the project will be calculated. 
	\item After all these computations, the requirements will be prioritized using the ratings provided by the stakeholders and project influences calculated from role and stakeholder influence. 
\end{itemize}
To solve the classic rank reversal problem, the following steps using prediction techniques are used. These steps are stated as follows:

\begin{itemize}
	\item When new requirements arrive, ratings given by a portion of stakeholders are elicited for each requirement.
	\item Item-to-Item collaborative filtering is then used to find similarity among already elicited requirements and the new requirements. Although collaborative filtering technique was used in \cite{lim2012stakerare}, they used it to find similar stakeholders instead of requirements. However, in this scenario it is more reasonable to find similar requirements first and then determine which stakeholders are more likely to rate those. Thus item-to-item technique was used.
	\item Model Based Collaborative Filtering using latent factors - learning parameter of users and feature vector of requirements are finally used for actual prediction of the values for the determined users from the previous step. This step also uses merged rating matrix and corresponding user-requirement relation matrix.
	\item Finally, StakeRare \cite{lim2012stakerare} is applied to the updated requirements list and new prioritized requirements list is attained from the approach.
\end{itemize}

\begin{figure}
	\centering
	\includegraphics[width=\columnwidth]{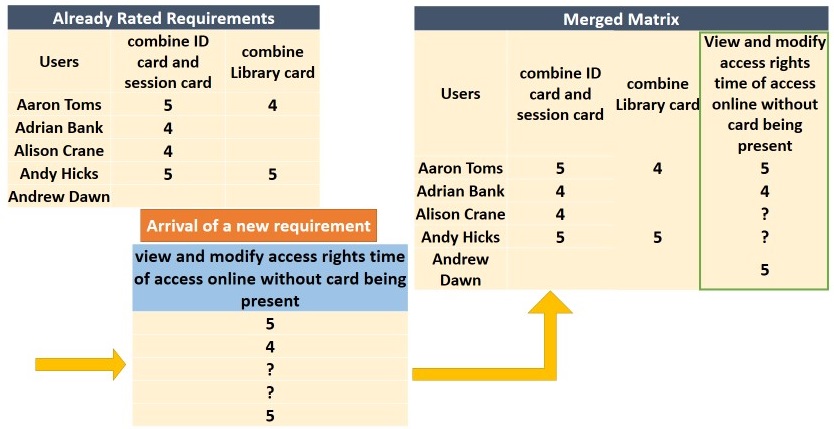}
	\caption{Merging rating matrices when a new requirement is added in the project}
	\label{fig_sim}
\end{figure}

A portion of the ratings are predicted rather than collecting all the ratings from stakeholders. Collecting all ratings from stakeholders for new requirements is time consuming and has scalability issues. Also large scale project developing process continues for several years. So it is natural for the stakeholders to provide appropriate rating after few years have passed on that project. Applying predicted ratings of new requirements eliminates these two problems. Lastly, the predicted ratings are used for prioritizing those newly arrived requirements.

\subsection{StakeRare}
The first four steps of the framework follows the StakeRare approach proposed in \cite{lim2012stakerare}. StakeRare at first prioritizes the stakeholders using social networks. Then it uses collaborative filtering to recommend requirements to relevant users. Then the requirements are prioritized based on importance of the role of the stakeholders on the project, importance of that stakeholder in that particular role and his/her actual rating given on the requirement \cite{lim2011social}. Finally the importance derived from project influence and rating which substitutes the actual rating of any user to any given requirement will be calculated and prioritization of the requirements will be made based on that. These steps can be completed by using Equations \ref{eq1} to \ref{eq4}.
{
\footnotesize
\begin{equation}
\label{eq1}
Influence_{role(i)} = \frac{RRmax+1-rank(role(i))}{\sum_{j=1}^{n}(RRmax+1-rank(role(j)))}
\end{equation}
\begin{equation}
\label{eq2}
	Influence_i = \frac{RSmax+1-rank(i)}{\sum_{k=1}^{n}(RSmax+1-rank(k))}
\end{equation}
\begin{equation}
\label{eq3}
	ProjectInfluence_i = Influence_{role(i)} * Influence_i
\end{equation}
\begin{equation}
\label{eq4}
	Importance_i = \sum_{i=1}^{n}(ProjectInfluence_i * r_i)
\end{equation}
}
Equation \ref{eq1} is used for prioritizing role. Here, RRmax is the maximum rank of the roles, rank(role(i)) is the rank of that role. Then Equation \ref{eq2} is used for calculating the influence of stakeholder in that role. Here, RSmax is the maximum rank of stakeholders in that role and rank(i) is the rank of that stakeholder. The influence of that stakeholder in that project is calculated using Equation \ref{eq3} which multiplies the stakeholder's influence in that role and influence of that role in the project. Then the importance of that requirement is calculated using summing all the ratings provided by the stakeholders in Equation \ref{eq4}.

\subsection{Stakeholder rating collection for new requirements}
As the framework adopts a semi-automated approach, manual collection of ratings from a portion of stakeholders needs to be done. Since ratings are collected from a subset of stakeholders for a newly arrived requirement, this framework decreases the number of human interactions necessary for updating prioritized requirements list. This approach will make the whole process more scalable for large scale requirements. After collecting the ratings derivation of user requirement relation matrix is generated for the new requirements.
Ratings are collected from stakeholders of different roles. Initially ratings are collected from on an average $M$ no. of stakeholders and this is decided based on extensive analysis. For new requirements on an average $K$ ratings ($K<M$) are collected from stakeholders. This ensures that no. of human interactions required for new requirements are always less than no. of human interactions necessary for previously elicited requirements. This step can quantify the reduction of human collaboration in requirements prioritization.

\subsection{Merging respective matrices}
Merging respective matrices is concerned with: merging of previous rating matrix and new rating matrix, merging of previous user-requirement relation matrix and user-requirement relation matrix for new requirements. User-requirement relation matrix is used in the next step to determine the probability of stakeholders to provide rating to lately considered requirements. Merged matrix is used to conduct actual prediction of the ratings. Figure \ref{fig_sim} illustrates the merging process.

User-requirement relation matrix consists of binary values: 1 and 0. If user-requirement relation matrix\textsubscript{(i,j)} = 1, it denotes that for i-th requirement specified j-th stakeholder has provided a rating and If user-requirement relation matrix\textsubscript{(i,j)} = 0, it denotes that for i-th requirement specified j-th stakeholder has not provided any rating. This matrix aids to find the pattern of stakeholders giving ratings to particular requirements. A sample of an user-requirement relation matrix is presented in Table \ref{tbl_RelationMatrix}.

\begin{table}[h]
	\centering
	\resizebox{.9\textwidth}{!}{\begin{minipage}{\textwidth}
	\caption{A Sample User-Requirement Relation Matrix}
	\label{tbl_RelationMatrix}
		\begin{tabular}{c|c|c|c}
			\hline
			Users        & \begin{tabular}[c]{@{}c@{}}a.3.1. combine ID \\ card and \\ session card\end{tabular} & \begin{tabular}[c]{@{}c@{}}a.3.2. combine \\ library card\end{tabular} & \begin{tabular}[c]{@{}c@{}}d.5.1. view and modify \\ access rights, time\\  of access, online, without \\ card being present\end{tabular} \\ \hline
			Aaron Toms   & 1                                                                                     & 1                                                                      & 1                                                                                                                                         \\ \hline
			Adrian Bank  & 1                                                                                     & 0                                                                      & 1                                                                                                                                         \\ \hline
			Alison Crane & 1                                                                                     & 0                                                                      & ?                                                                                                                                         \\ \hline
			Andy Hicks   & 1                                                                                     & 1                                                                      & ?                                                                                                                                         \\ \hline
			Andrew Dawn  & 0                                                                                     & 0                                                                      & 1                                                                                                                                         \\ \hline
		\end{tabular}
	\end{minipage} }
\end{table}

\subsection{Determining users for prediction}
Ratings of which users are to be predicted has to be determined first to implement actual prediction. To determine, similarity among new and previously elicited requirements can be used. Item to item collaborative filtering can be used for finding similarities among requirements. This approach predicts probability of users to give ratings to new requirements based on this similarities. It learns if any specified user tends to rate the new requirement based on his/her rating on similar previous requirements. This approach can be divided into two steps. The steps are stated as following:

\subsubsection{Correlation Computation} For item based collaborative filtering to work, similarity among items has to be figured. In this case, requirements are the items and similarity among these requirements are estimated using correlation analysis. There are three correlation techniques which were considered for finding correlation among requirements. The techniques - Pearson Coefficient Correlation, Cosine Similarity and Jaccard Distance.

All of these techniques are implemented on the user-requirement relation matrix\textsubscript{(p,q)}  to detect similarities among requirements. Best results are produced by cosine similarity and Pearson coefficient correlation as both of these approaches are invariant to scaling. This means similarities among elements are invariant even if all elements are multiplied by a nonzero constant. However, cosine similarity is not invariant when any constant is added to all elements. But Pearson correlation is also invariant to adding any constant to all elements. For example, if there are two vectors X1 and X2, and Pearson correlation function is called pearson(), pearson(X1, X2) == pearson(X1, 2 * X2 + 3). This property is really important as we are looking for similarity patterns among items. The items do not need to be exactly identical to be affirmed similar by our approach. Hence, Pearson Coefficient Correlation is used to determine similarity among items and used to predict ratings of the stakeholders. The equation of Pearson Coefficient Correlation is stated in Equation \ref{eq5}.

{\footnotesize
\begin{equation}
\label{eq5}
sim(i,j) = \frac{\sum_{u\in U}^{}(R_{(u,i)}-\bar{R_i}) (R_{(u,j)}-\bar{R_j})}{\sqrt{\sum_{u\in U}^{}(R_{(u,i)}-\bar{R_i})^2)} \sqrt{\sum_{u\in U}^{}(R_{(u,j)}-\bar{R_j})^2)}}
\end{equation}
}
	
To ensure accuracy of the correlation computation, we must first isolate the co-rated cases (i.e., cases where the users rated both i and j items). Let the set of users who both rated i and j are denoted by U then the correlation similarity is given by Here R\textsubscript{u,i} denotes the rating of user u on item i, $\overline{R_i}$ is the average rating of the i-th item. Hereafter, using this similarity function an Requirement-to-Requirement similarity matrix, as presented in Table \ref{tbl_sim}, will be generated.

\begin{table}[h]
\caption{A Sample Requirement-to-Requirement Similarity Matrix}
\label{tbl_sim}
\centering
\begin{tabular}{c|c|c|c}
\hline
  & Requirement 1 & Requirement 2 & Requirement 3\\
\hline
Requirement 1 & 1 & 0.76 & 0.78\\
\hline
Requirement 2 & 0.76 & 1 & 0.86\\
\hline
Requirement 3 & 0.78 & 0.86 & 1\\
\hline
\end{tabular}
\end{table}

\subsubsection{Stakeholder Selection} Using the similarity matrix obtained from the previous step, stakeholders likely to rate a requirement can be predicted based on the commonly used Equation \ref{eq6}.
{
\footnotesize
\begin{equation}
\label{eq6}
P_{(u,i)} = \frac{\sum_{all similar items,N}^{} (S_{i,N} * R_{u,N})}{\sum_{all similar items,N}^{} |S_{i,N}|}
\end{equation}
}
By using weighted sum we can predict the value for any user-item pair. First we take all the items similar to our target item, and from those similar items, we pick items which the active user has rated which is denoted by $S_{i,N}$. The actual rating given by the user U is denoted as $R_{u,N}$ in the equation. We weight the user's rating for each of these items by the similarity between that and the target item. Finally, we scale the prediction by the sum of similarities to get a reasonable value for the predicted rating. For user u and item i Predicted rating is denoted as $P_{u,i}$. These predicted values are used for calculating actual predicted rating by the users. These values can be used to suggest requirements to a user.	

\subsection{Prediction of ratings}
For predicting the value of a rating from a particular user, Model based collaborative filtering is used. The benefit of such technique is that it considers latent factors \cite{agarwal2009regression}. These factors are not explicitly stated rather than inferred based on the statistical analysis of any specified scenario. There are two latent factors are related to prediction in the scenario illustrated in the paper. For each user, we have to calculate the learning parameter ($\theta$) and each requirement is associated with a feature vector (x). 
For each of the stakeholders learning parameters and for each of the requirements feature vector is initialized to small random values primarily. A cost function $J$ \cite{rosasco2004loss} using those two factors is minimized to obtain actual learning parameters and feature vectors. Minimization of those parameters are completed using gradient descent \cite{burges2005learning} technique. Finally, the predictions of ratings are made by using multiplication of transpose matrix of learning parameter and the matrix derived from feature vector. 
The methodology is presented below:

\begin{itemize}
	\item For each user j we have to learn the parameter $\theta^{(j)} \in R^{(n+1)}$ where n= number of features for predicting the ratings of new requirements. It denotes that $\theta^{(j)}$ is a vector which has n+1 dimensions. Given the feature vector $x^{(i)}$ for $i^{th}$ requirement using linear regression modeling we can formulize the problem of deducing parameter vector.
	\item For every requirement i we have to learn the feature vector $x^{(i)} \in R^{(n+1)}$ where n = number of features for predicting the ratings of new requirements. It denotes that  $x^{(i)}$ is a vector which has n+1 dimensions. Given the parameter vector $\theta^{(j)}$ and actual rating  $y^{(i,j)}$ for $j^{th}$ stakeholder we can formulize the problem of inferring feature vector using linear regression modeling.
	\item It should be noted that, parameter vector $\theta$ and feature vector $x$ both should be initialized to small random values for initial computation. Then the cost function $J$ is used to estimate and adjust the values of $\theta$ and $x$ simultaneously to fulfill the objective of minimization. Henceforth, parameter vector $\theta$ and feature vector x is derived for each of the requirements and stakeholders. 
	Based on these, the prediction value can be calculated by Equation \ref{eq7}. It means that, for $i^{th}$ requirement and $j^{th}$ user the predicted value is  $\theta^{(j)}$ transposes $x^{(i)}$.
	{\footnotesize
		\begin{equation}
		\label{eq7}
		(\theta^{(j)})^T x^{(i)}
		\end{equation}
	}
\end{itemize}

\section{Experimental Setup}
\label{sec:exp}
SAFFRON uses StakeRare \cite{lim2012stakerare} for prioritizing requirements and stakeholders. Thus, a prototype of StakeRare was implemented using JAVA. To implement the framework real life datasets was required. In addition, for evaluating the efficiency and effectiveness of the framework, right research questions need to be set. A brief discussion about these procedures are presented in the following sub-sections.

\subsection{RALIC dataset}
The RALIC  \cite{lim2012stakerare} project was used for implementation and experimentation of the proposed approach. The full form of RALIC is Replacement Access, Library and ID Card project. It was a software project which was developed to maintain the access control system at University College London (UCL). The main reason for selecting this project was that it is complete and reliable. Besides another criterion was its scale. RALIC project had a complex stakeholder base, where there are more than 60 groups and 30,000 system users. These stakeholders have different and sometimes conflicting requirements.The dataset has more than 3,000 ratings from the stakeholders. For our experimentation, 82 requirements and 62 stakeholders are selected from RALIC dataset. As only one complete and reliable dataset is used, repeated random sub-sampling is implemented in order to eliminate skewed behavior of the dataset. Repeated iterations are applied 30 times for each experimentation setting.

\subsection{Research questions}
The main goal of SAFFRON is to predict ratings of stakeholders for requirements. To fulfill this goal experimentation has been performed on various experimental setting. Finding best experimentation setting depends on following research questions-
\begin{itemize}
	\item \textbf{RQ1:} How many previously elicited and rated requirements are sufficient to predict missing values?
	\item \textbf{RQ2:} Ratings of how many stakeholders on an average for requirements are enough to predict the rating for missing values of new requirements? 
	\item \textbf{RQ3:} What percentage of missing values should be predicted to ensure consistency of prioritization process?
\end{itemize}
Besides, to measure the performance of the framework and ensure its effectiveness following research questions must be addressed:
\begin{itemize}
	\item \textbf{RQ4:} What is the correlation of SAFFRON to ground truth comparing to state of the art approach - StakeRare? 
	\item \textbf{RQ5:} What percentage of human interactions could be reduced by SAFFRON?
\end{itemize}

\begin{figure}[h]
	\centering
	\begin{multicols}{2}
		\includegraphics[width=\linewidth]{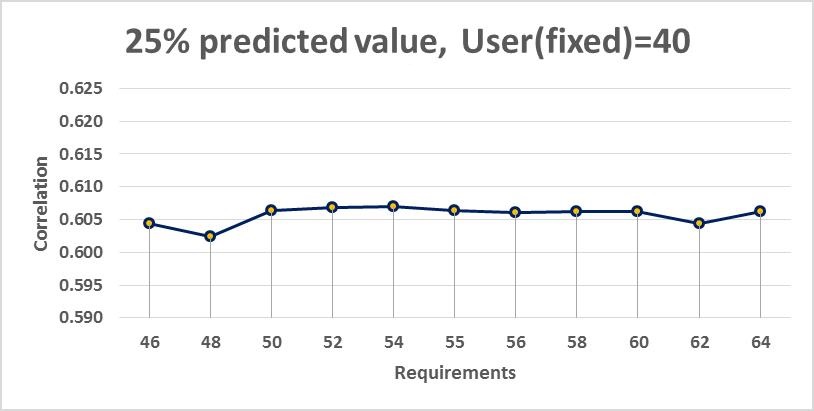}\par 
		\includegraphics[width=\linewidth]{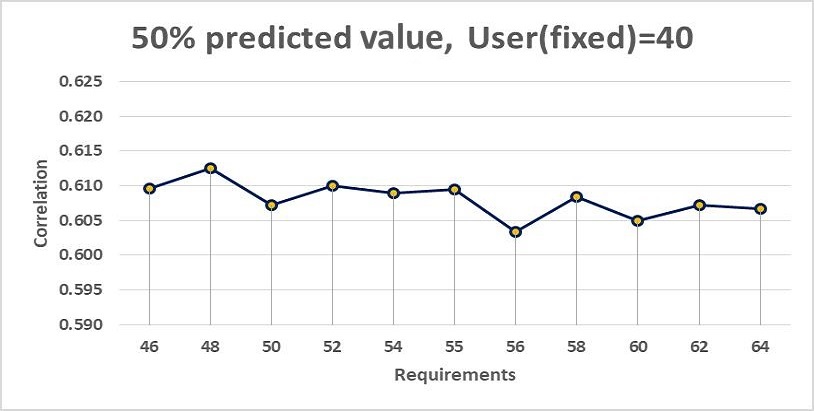}\par 
	\end{multicols}
	\begin{multicols}{2}
		\includegraphics[width=\linewidth]{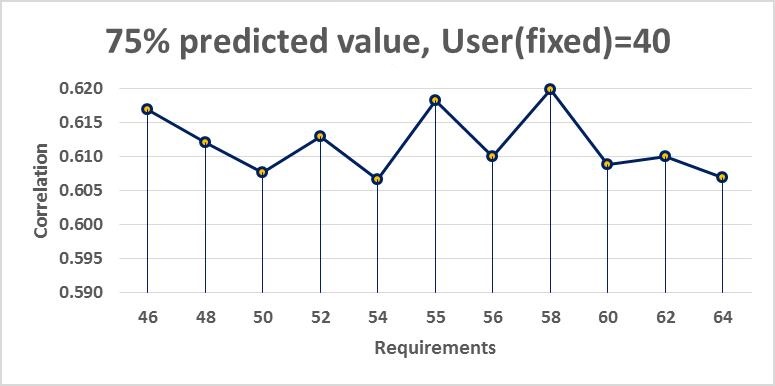}\par
		\includegraphics[width=\linewidth]{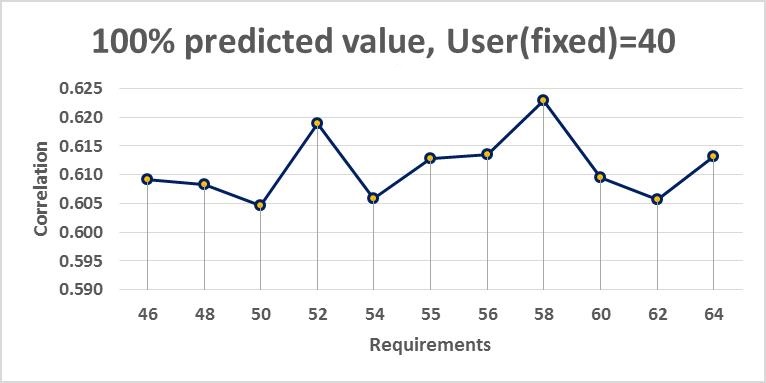}\par
	\end{multicols}
	\caption{Ranking Correlation with Ground Truth for Top 25\%, 50\%, 75\% and 100\% Predicted Rating for Varying No. of Requirements}
	\label{fig_graphRequirement}
\end{figure}

\begin{figure}[h]
	\centering
	\begin{multicols}{2}
		\includegraphics[width=\linewidth]{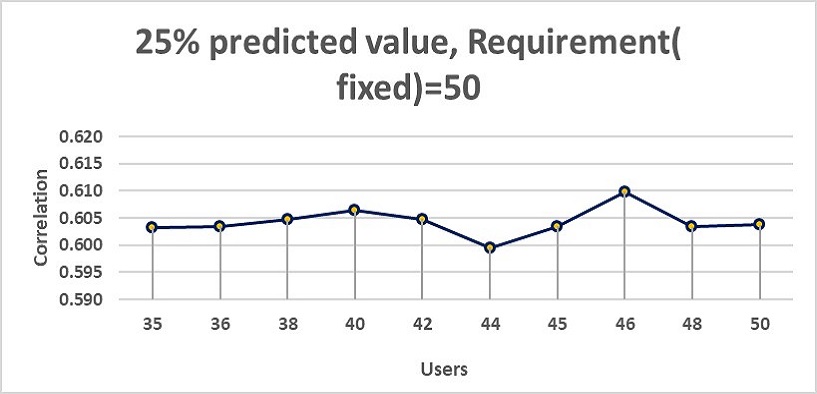}\par 
		\includegraphics[width=\linewidth]{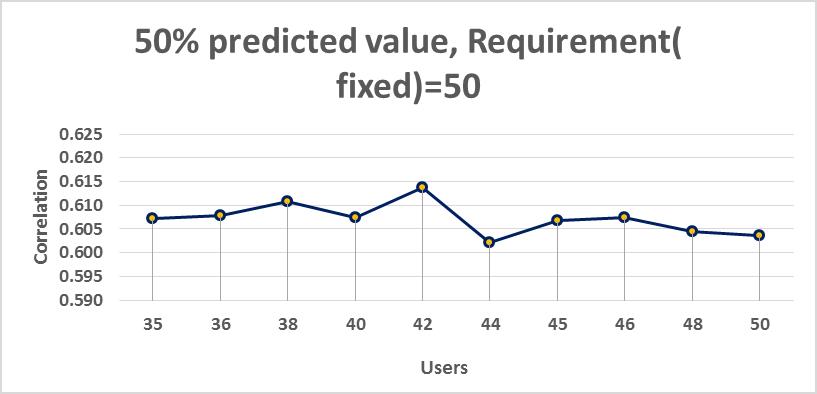}\par 
	\end{multicols}
	\begin{multicols}{2}
		\includegraphics[width=\linewidth]{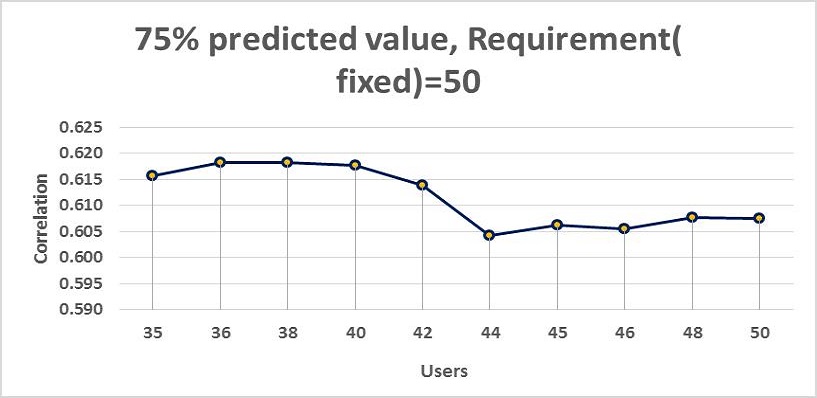}\par
		\includegraphics[width=\linewidth]{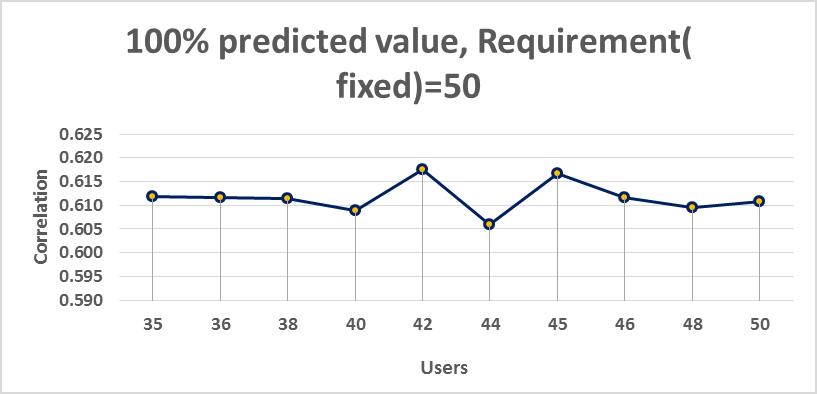}\par
	\end{multicols}
	\caption{Ranking Correlation with Ground Truth for Top 25\%, 50\%, 75\% and 100\% Predictions for Varying No. of Users}
	\label{fig_graphUser}
\end{figure}

\begin{figure}
	\begin{subfigure}{.5\textwidth}
		\centering
		\includegraphics[width=\linewidth]{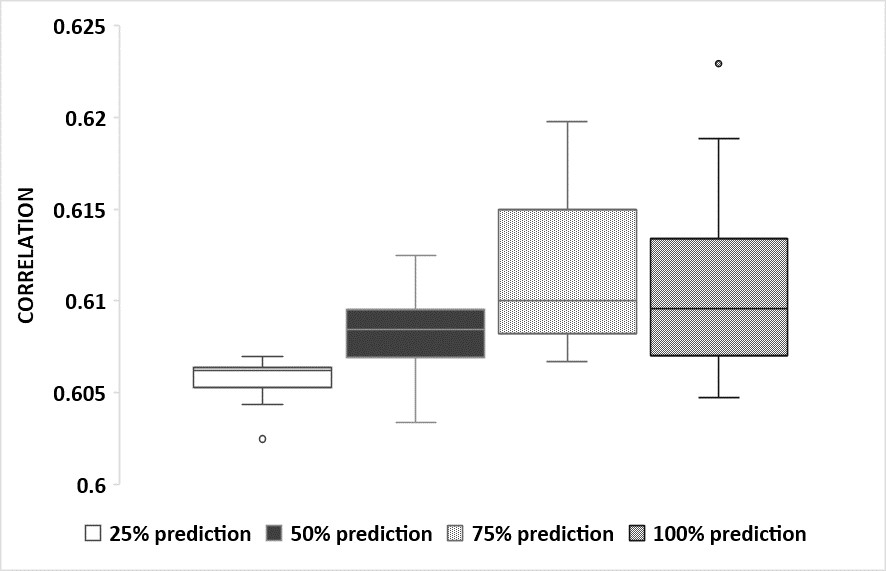}
		\caption{Ranking correlation when no. of training requirements are changing}
		\label{fig:sfig1}
	\end{subfigure}%
	\begin{subfigure}{.5\textwidth}
		\centering
		\includegraphics[width=\linewidth]{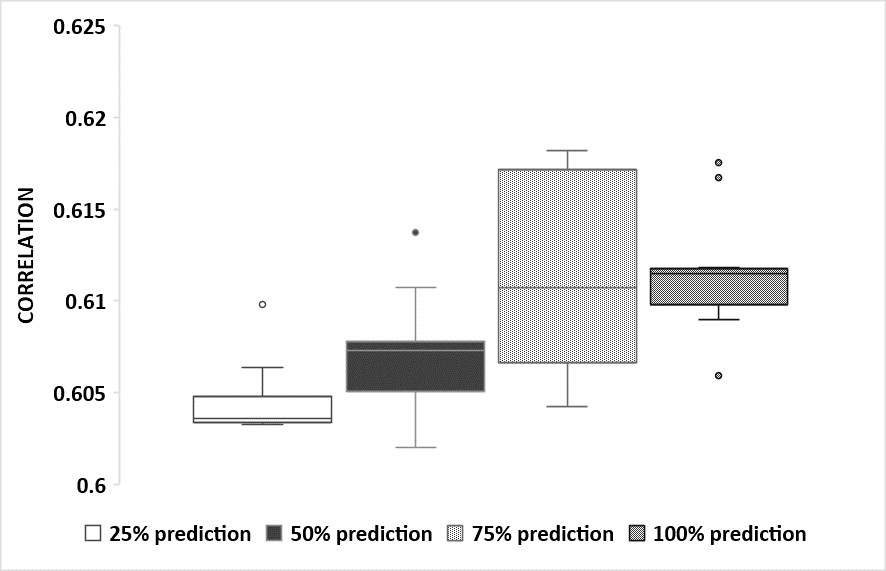}
		\caption{Ranking correlation when no. of users are changing}
		\label{fig:sfig2}
	\end{subfigure}
	\caption{Boxplots of ranking correlation of SAFFRON with ground truth}
	\label{fig_correlationVariability}
\end{figure}

Addressing these research questions will assist to accomplish the effectiveness of this software requirements prioritization framework in solving rank reversal problem and limiting human interaction.

\section{Result Analysis}
\label{sec:result}

The proposed framework was implemented by differing no. of previously elicited requirements, no. of stakeholders provided ratings on new requirements and percentage of top-N predicted values. Experimental settings of various combination of above mentioned parameters were tested. Spearman's ranking correlation was used to measure the ranking correlation among Ground Truth, StakeRare and SAFFRON. 

To address RQ1, no. of previously elicited requirements were changed whereas no. of users collaborated manually for new requirements remained fixed (user = 40). Then the correlation is calculated. From Figure \ref{fig_graphRequirement} it could be seen that if 46-58 requirements were elicited in initial stage from 82 requirements, better ranking correlation to the ground truth is exhibited for new requirements. The ranking correlation then usually goes down with the increase of training requirements. This is understandable as overfitting may occur due to using too many requirements.

RQ2 is concerned with no. of users giving rating to new requirements to accurately predict ratings for other users. So in our experiment, we kept the training requirements fixed (50) and varied the no. of users. From Figure \ref{fig_graphUser} it is seen that if ratings from 38-45 stakeholders are collected for new requirements among 62 stakeholders then enhanced ranking correlation is achieved after prediction. The correlation gradually decreases, if the no. of stakeholders giving rating to new requirements goes above 45.

Too much prediction can make a system perform inconsistently. So it is important to know what percentage of values should be predicted, which is the concern of RQ3. Figure \ref{fig_correlationVariability} illustrates that higher predictions can sometime result in high ranking correlation. However, the variance of ranking correlation is also high in that case. So, a better performance is always not guaranteed. On the other hand,  lower predictions have low ranking correlation on average. However, its variance is much lower, making it more consistent.

Based on the observations from RQ1-RQ3, we evaluated SAFFRON in different experimental settings. Here, we also varied the number of new requirements that are added later in the projects. This actually creates the rank reversal scenario. For those requirements, we considered that certain no. of ratings are given by the stakeholders. Rest of the ratings were predicted by SAFFRON and then the ranking correlation was computed. Results obtained from the experiment is presented in Table \ref{table_CorrelationGroundTruth}.

\begin{table}[h]
	\caption{Performance evaluation of SAFFRON in terms of ranking correlation and reduced human interaction}
	\label{table_CorrelationGroundTruth}
	\resizebox{.54\textwidth}{!}{\begin{minipage}{\textwidth}
			\centering
			\begin{tabular}{c|c|c|c|c|c|c}
				\hline
				\multirow{3}{*}{\textbf{Experimental Setting}} & \multirow{3}{*}{\textbf{\begin{tabular}[c]{@{}c@{}}No. of New \\ Requirement\end{tabular}}}	&\multicolumn{2}{|c|}{\textbf{StakeRare}}                       & \multicolumn{3}{|c}{\textbf{SAFFRON}}                                                                            \\ \cline{3-7}
				\textbf{} & \textbf{}                   & \textbf{Correlation} & \textbf{\begin{tabular}[c]{@{}c@{}}No. of Users \\ Communicated\end{tabular}} & \textbf{Correlation} & 
				\textbf{\begin{tabular}[c]{@{}c@{}}No. of Users \\ Communicated\end{tabular}} & \textbf{ \begin{tabular}[c]{@{}c@{}}\% of reduced human \\ interaction\end{tabular}} \\ \hline
				Req.=40, User=35 & 10           & 0.923699314          & 48                                    & 0.923140505          & 35                                    & 27\%                                             \\ \hline
				Req.=50, User=40 & 15           & 0.815667899          & 51                                    & 0.815205913          & 40                                    & 21.6\%                                           \\ \hline
				Req.=50, User=45 & 20           & 0.76699411           & 54                                    & 0.767627872          & 45                                    & 16.7\%                                             \\ \hline
				Req.=55, User=45 & 25          & 0.675217954          & 52                                    & 0.675609783          & 45                                    & 13.5\%                                             \\ \hline
				Req.=50, User=45 & 30           & 0.675217954          & 54                                    & 0.67317174           & 45                                    & 16.7\%                                             \\ \hline
			\end{tabular}
		\end{minipage} }
	\end{table}
	
	\begin{table}[h]
		\caption{RMSE for changing no. of requirements (users fixed = 40)}
		\label{tbl_RMSforFixedUser}
		\resizebox{.66\textwidth}{!}{\begin{minipage}{\textwidth}
				\centering
				\begin{tabular}{ccccc}
					\hline
					\textbf{\begin{tabular}[c]{@{}c@{}}Experimental\\ Setting\end{tabular}}                                 & \textbf{\begin{tabular}[c]{@{}c@{}}RMSE for \\ 25\% Prediction\end{tabular}} & \textbf{\begin{tabular}[c]{@{}c@{}}RMSE for \\ 50\% Prediction\end{tabular}} & \textbf{\begin{tabular}[c]{@{}c@{}}RMSE for\\ 75\% Prediction\end{tabular}} & \textbf{\begin{tabular}[c]{@{}c@{}}RMSE for \\ 100\% Prediction\end{tabular}} \\ \hline
					\textbf{\begin{tabular}[c]{@{}c@{}}Requirement (Train) = 50\\ User (Manual Rating) = 40\end{tabular}} & 0.000939000                                                                  & 0.001882290                                                                  & 0.002364495                                                                 & 0.003137046                                                                   \\ \hline
					\textbf{\begin{tabular}[c]{@{}c@{}}Requirement (Train) = 55\\ User (Manual Rating) = 40\end{tabular}} & 0.000863000                                                                  & 0.001661692                                                                  & 0.002521551                                                                 & 0.002894824                                                                   \\ \hline
					\textbf{\begin{tabular}[c]{@{}c@{}}Requirement (Train) = 60\\ User (Manual Rating) = 40\end{tabular}} & 0.000809000                                                                  & 0.001336428                                                                  & 0.001833125                                                                 & 0.002372170                                                                   \\ \hline
				\end{tabular}
			\end{minipage} }
		\end{table}
		
		\begin{table}[h]
			\caption{RMSE for changing no. of users (requirements fixed = 50)}
			\label{tbl_RMSforFixedRequirement}
			\resizebox{.66\textwidth}{!}{\begin{minipage}{\textwidth}
					\centering
					\begin{tabular}{ccccc}
						\hline
						\textbf{\begin{tabular}[c]{@{}c@{}}Experimental\\ Setting\end{tabular}}        &
						\textbf{\begin{tabular}[c]{@{}c@{}}RMSE for \\ 25\% Prediction\end{tabular}} & \textbf{\begin{tabular}[c]{@{}c@{}}RMSE for \\ 50\% Prediction\end{tabular}} & \textbf{\begin{tabular}[c]{@{}c@{}}RMSE for\\ 75\% Prediction\end{tabular}} & \textbf{\begin{tabular}[c]{@{}c@{}}RMSE for \\ 100\% Prediction\end{tabular}} \\ \hline		
						\textbf{\begin{tabular}[c]{@{}c@{}}Requirement (Train) = 50\\ User (Manual Rating) = 40\end{tabular}} &
						0.000939000              & 0.001882290              & 0.002364495              & 0.003137046               \\ \hline
						\textbf{\begin{tabular}[c]{@{}c@{}}Requirement (Train) = 50\\ User (Manual Rating) = 45\end{tabular}} &
						0.000893000              & 0.001382774              & 0.002100870              & 0.002904053               \\ \hline
						\textbf{\begin{tabular}[c]{@{}c@{}}Requirement (Train) = 50\\ User (Manual Rating) = 50\end{tabular}} &
						0.000687000              & 0.000999000              & 0.001428481              & 0.002181204               \\ \hline
					\end{tabular}
				\end{minipage} }
			\end{table}

From Table \ref{table_CorrelationGroundTruth} it is seen that SAFFRON and StakeRare have almost similar ranking correlation with the Ground Truth. This answers our RQ4 that SAFFRON is as effective as StakeRare. It can be said that SAFFRON can also solve the rank reversal problem. Another significant finding is that SAFFRON reduces human interaction in all cases. Human interaction were lessened from StakeRare approach by 27\%, 21.6\%, 16.7\%, 13.5\% and 16.7\% respectively in the 5 experimental settings presented in Table \ref{table_CorrelationGroundTruth}. So in a nutshell 13.5-27\% human interaction is reduced, which also answers RQ5.

After prediction of ratings by SAFFRON, the missing values of the selected part in the requirement-stakeholder matrix is filled with calculated predicted ratings. We compared the selected part of the updated matrix with the original ratings of that fragment derived from Ground Truth. Deviation of predicted ratings and original ratings was measured using Root Mean Squared Error (RMSE). This actually provided the rationale behind the performance of our approach.

A smaller RMSE indicates that predicted ratings are more closer to the original ratings. Table \ref{tbl_RMSforFixedUser} and \ref{tbl_RMSforFixedRequirement} present RMSE for various experimental settings. It can be seen that the RMSE values are not significant which resulted in SAFFRON's better performance. It is also observed that for 25\% prediction value, the value of RMSE is lowest. This is another reason for which lower number of predictions can be used.

\section{Conclusion and Future Work}
\label{sec:conc}
This paper proposes a framework that addresses rank reversal problem in software requirements prioritization and reduces the no. of human interaction in the process. It used item based collaborative filtering to find similarity among previously rated requirements and newly arrived requirements. By using those similarities among requirements, probability of users to rate new requirements are computed. Ratings are then predicted, for users having high probabilities, adopting model based collaborative filtering. More precisely, regression techniques utilizing gradient descent to minimize cost function of latent factors is used for predicting ratings. Results suggests that the framework reduces human interaction while updating prioritized requirements list and also maintains consistent ranking correlation with ground truth compared to state of the art approaches.

One of the future challenges of the work is to cluster requirements and stakeholders based on prior information. Clustering can aid to find patterns from already elicited stakeholder ratings. Prediction will be more accurate and effective if the collaborative algorithm is applied on clustered requirements and stakeholders. Therefore, there is a scope of improvement by extending the framework by using clustering techniques.
\section*{Acknowledgment}
This research is funded by the fellowship program from Information and Communications Technology (ICT) division of Government of People's Republic of Bangladesh. The award no. is 56.00.0000.028.33.073.16-50.



\bibliographystyle{IEEEtran}
%

%

\end{document}